# Evaluating Performance Consistency in Competitive Programming: Educational Implications and Contest Design Insights

Zhongtang Luo and Ethan Dickey
*Department of Computer Science*
*Purdue University*
West Lafayette, USA
luo401@purdue.edu, dickeye@purdue.edu

**Abstract**—Competitive programming (CP) contests are often treated as interchangeable proxies for algorithmic skill, yet the extent to which results at lower contest tiers anticipate performance at higher tiers, and how closely any tier resembles the ubiquitous online-contest circuit, remains unclear. We analyze ten years (2015–2024) of International Collegiate Programming Contest (ICPC) standings, comprising five long-running superregional championships (Africa & Arab, Asia East, Asia West, North America, and Northern Eurasia), associated local regionals of North America and Northern Eurasia, and the World Finals. For 366 World Finalist teams (2021–2024) we augment the dataset with pre-contest Codeforces ratings. Pairwise rank alignment is measured with Kendall's $\tau$.

Overall, superregional ranks predict World Final ranks only moderately (weighted $\tau = 0.407$), but regional-to-superregional consistency varies widely: Northern Eurasia exhibits the strongest alignment ($\tau = 0.521$) while Asia West exhibits the weakest ($\tau = 0.188$). Internal consistency within a region can exceed its predictive value for Worlds — e.g., Northern Eurasia and North America regionals vs. superregionals ($\tau = 0.666$ and $\tau = 0.577$, respectively). Codeforces ratings correlate more strongly with World Final results ($\tau = 0.596$) than any single ICPC tier, suggesting that high-frequency online contests capture decisive skill factors that many superregional sets miss.

We argue that contest organizers can improve both fairness and pedagogical value by aligning problem style and selection rules with the formats that demonstrably differentiate teams, in particular the Northern-Eurasian model and well-curated online rounds. All data, scripts, and additional analyses are publicly released to facilitate replication and further study.

## 1 Introduction

Competitive programming (CP) has become widely recognized as an effective educational tool in computer science, primarily for its role in enhancing students' problem-solving skills, computational thinking, and algorithmic proficiency. Empirical studies consistently highlight CP's educational benefits, such as improved student proactivity, reduced perceived difficulty of programming concepts, and increased retention rates in introductory programming courses [1], [4]. Additionally, programming contests promote independent learning, stimulate innovative thinking, and actively engage students in complex computational problem-solving [27]. Garcia and Aguirre (2014) provide further empirical support by demonstrating measurable skill progression through sustained participation in competitive programming [8]. Finally, Yuen et al. (2023) also emphasize CP's practical educational impact, noting that contests stimulate student interest and enhance independent learning, innovative thinking, and problem-solving skills, which benefits student employability and preparedness for advanced computational challenges [35].

Several educational frameworks offer robust theoretical justifications for incorporating CP into curricula. From a constructivist learning theory perspective, CP facilitates active learning through iterative experimentation, rapid feedback, and collaborative problem-solving experiences [23], [34]. Competitive learning frameworks further advocate for CP's structured competitive interactions, effectively supporting individualized learning alongside cooperative team dynamics [4], [13], [17], [21]. Motivational theories, such as Self-Determination Theory, help illuminate how CP contests intrinsically motivate students by meeting psychological needs for autonomy, competence, and relatedness, particularly when combined with gamification elements [10], [28]. Furthermore, Gonzalez-Escribano et al. (2019) illustrate how competitive environments can be intentionally structured to encourage collaborative behaviors, creating a balanced educational environment that leverages both competition and cooperation [9]. For further review of competitive programming and its place in computer science education, please refer to [15], [35].

Practically, there have been several studies on effective teaching of and participation in CP courses and contests. Mascio et al. study (further) gamification of CP through awarding badges and providing a tailored "next problem" recommendation system [6]. The tool they created, *Oii-web*, nearly doubled participation in the CP contest Olympiads in Informatics (OII) for secondary school students in Italy [5]. In a similar vein, others have studied course design with some experience reports showing increased engagement and academic student success [20], [16], [7], provided contest strategy guides for participation in CP contests [2], [18], [32], and have shown that introducing CP-style assignments and tools into traditional courses, such as a CS1 course taught in Haskell, increases amount of programming done by students as well as students' diligence, rigor, and joy in programming [29].

Despite this substantial empirical evidence and strong theoretical support for CP as a subset of competitive learning for computer science [30], there remains a gap in understanding how the structure and design of competitive programming events influence educational outcomes and student performance. Some studies have analyzed trends in participation and provided recommendations for increasing participation in North American superregional contests [3], but analysis of structural differences between contests has not been explored. Specifically, preeminent contests like the International Collegiate Programming Contest (ICPC) exhibit substantial variability in superregional contest rules and qualifications, problem topics, and problem difficulties, yet the implications of these structural differences have not been thoroughly explored. Moreover, educators and organizers frequently rely on anecdotal evidence, experience, and intuition rather than empirical insights when designing and adapting CP contests.

This study seeks to shed some light on these issues by empirically examining team performance consistency across various ICPC contests. Rather than focusing explicitly on detailed problem-setting practices — which remain opaque in most cases — we analyze correlations among performance metrics at different competitive levels. Additionally, we investigate the predictive validity of online competitive programming ratings, such as those from platforms like Codeforces, hypothesizing that their frequent and diverse contest formats provide reliable indicators of competitor skill and potential contest outcomes.

This research is guided by three primary research questions:

RQ1  Does team performance consistency differ significantly among ICPC superregional contests?
RQ2  How accurately do online competitive programming ratings predict ICPC contest outcomes, and do these ratings align differently with particular superregional contests?
RQ3  What insights can an analysis of performance consistency provide to help educators and contest organizers improve the design and effectiveness of competitive programming events?

By investigating these RQs, this study aims to inform evidence-based improvements in contest design, enhancing the alignment of competitive programming events with educational theories and practical learning outcomes and ultimately promoting more effective educational practices in computer science education.

## 1.1 ICPC Overview

The International Collegiate Programming Contest (ICPC) is an annual, multi-tiered, worldwide, collegiate competitive programming competition. It is generally regarded as the oldest (1970) and most prestigious programming contest in the world. In this contest format, universities send teams of three students to compete, each of which is tasked to solve 10–13 algorithmic problems within five hours under the restraints of one computer per team, no internet access, and one 25-page document of reference material. A brief, readable history of ICPC can be found in [14]. A much more detailed one can be found on Wikipedia[1].

ICPC contests are divided into multiple levels. At the top level, the annual World Finals (WF) host the most competitive teams from around the world to compete in one final contest. The lower levels are divided into eight regions: Africa and Arab, Asia East, Asia Pacific, Asia West, Europe, Latin America, North America, and Northern Eurasia. World Finals chooses a number of teams that can be sent from each region. Each region then sets its own rules for promoting competitive teams from superregional finals to World Finals. We give an overview of the structure in Figure 1.

## 1.2 ICPC Superregional Structure and Qualifications

We provide a concise overview of each region's structure and selection of World Finals teams as of Spring 2025. Using this information, we select 2 regions to analyze and compare regional-to-superregional consistencies in Sections 3.3 and 3.4: the North America and Northern Eurasia regions.

**Africa & Arab (ACPC)**
- **WF qualification:** *Only* through the Africa & Arab Collegiate Programming Championship (ACPC); no direct promotion from country-level regionals.
- **ACPC entry:** One university-level local contest → one country-level regional → ACPC. Country quotas are set annually by ACPC staff, primarily by historical participation.
- **Judging:** ACPC Scientific Committee invites experienced coaches, ex-contestants, prior judges, and professionals familiar with ICPC; selection is usually based on recommendations and past involvement.

**Asia East (AECF)**
- **WF qualification:** The winning teams of every invitational regional *and* the top teams at the East Continent Final advance to the World Finals; the East Continent Final host university also receives an automatic slot.
- **East Continent Final entry:** Invitational. Slots are awarded to regional medalists, regional host schools, universities that reached the World Finals in the previous three years, and other schools selected by ICPC Beijing Headquarter [11].
- **Judging:** Each regional or superregional host recruits its own authors and judges; commercial problem-setting services may be contracted [31].

**Asia Pacific (APAC)**

---

1. https://en.wikipedia.org/wiki/International_Collegiate_Programming_Contest

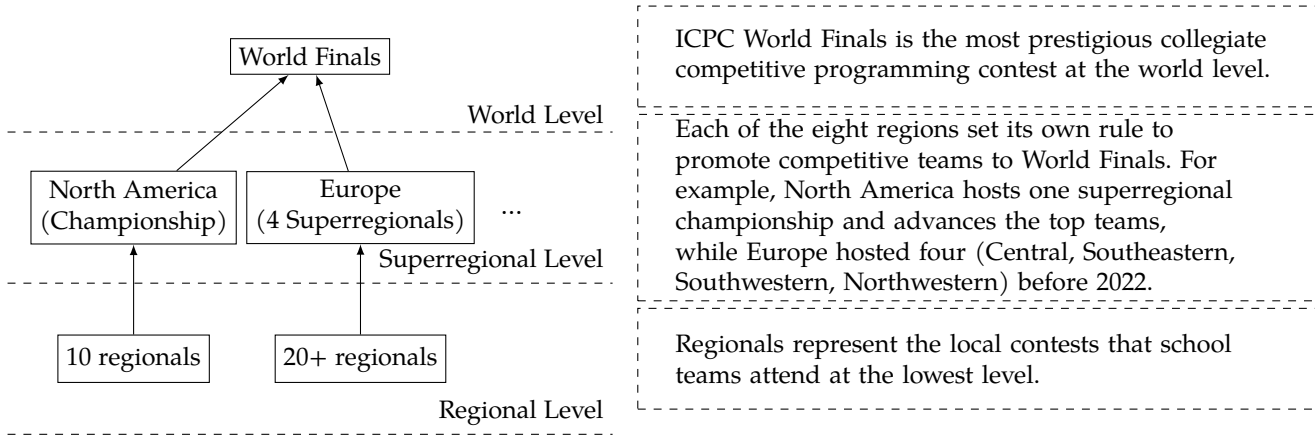

Fig. 1. Hierarchy of ICPC contests.

- **WF qualification:** (*2024-25 cycle*[2]) Each APAC regional winner (after removing South Pacific and other superregions) earns an automatic World Finals slot. If the same university wins multiple regionals, it may advance only one team, chosen by the school or via the Championship (rules B1-B3). Remaining Asia-Pacific slots are filled by the highest-ranked universities in the Asia Pacific Championship after filters F1(1)-F1(2).
- **Championship entry:**
  - Automatic invites - all regional winners (D1) and the top two teams from the South Pacific Independent Regional (D2).
  - Ranked list - the remaining slots come from a merged cross-regional list ordered by $(R-1)/S$ after filters D3(1)-D3(5); capped at three teams per university, guaranteeing at least one team per APAC country, plus up to one wildcard selected by the director (D4(2-5), D5).
  - Each team may compete in at most two regionals per season (A1).
- **Judging:** Each regional nominates scientific committee members; nominees elect a Chief Judge (last two cycles: Jonathan I. Gunawan, SGP). Final panel spans 6-8 Asia-Pacific countries.

**Latin America (LAC)**
- **WF qualification:** Six regional winners + next best overall teams at the Latin-America Championship (Programadores de América). No direct WF slots from the first-stage regional.
- **Championship entry:** Fixed regional quotas, plus inclusivity extras (female-only teams, countries otherwise unrepresented). Slots are earned only via the single synchronized first-stage contest [19].
- **Judging:** Open call for problem proposals; Latin America Chief Judge chooses judges and setters to form a balanced panel.

**North America (NAC)**
- **WF qualification:** Top $N(\approx 17)$ teams from the North America Championship (NAC), one per university.
- **NAC entry:** Eleven regional contests; 50 teams advance from regionals, each regional receives a quota proportional to historical strength and participation. Only the highest-ranked team per school advances.
- **Judging:** NAC Chief and Deputy Judges invite problem setters and judges from across the region.

**Northern Eurasia (NEF)**
- **WF qualification:** Top $N(\approx 16)$ teams from the Northern Eurasia Finals (NEF), one per university; extra host-country slots possible.
- **NEF entry:** Seventeen regional contests receive dynamic quotas computed by a published formula[3] (affected by NEF workstation limit, past performance, number of teams participating in first rounds, etc.).
- **Judging:** Mailing-list call to an approved pool of former judges/authors; Chief Judge curates the final problem set and adds people to the listserv.

**Europe (EUC)**
- *Details pending from regional contacts*[4]. *Provisional note: 2024 introduced a single Europe Championship fed by four long-standing superregionals (NW, SE, SW, Central).*

**Asia West (AWCF)**
- *Details pending from regional contacts.*

### 1.3 Population Sample

Five of the ICPC regions have consistently named superregional championships that promote their teams to World Finals that extend beyond one year. We list them in Table 1, and use them as the dataset for our comparison against the World Finals and Codeforces. Northern Eurasia was listed as Northeastern Europe Regional Contest from 2015–2017 but functionally acted as its own region in structure and promotion to worlds, so 2015–2017 data is included in the analysis.

Additionally, as the Europe Championship was added in 2024, and the regionals (Northwest, Southeast, Southwest, Central) have been organized as effectively superregionals pre-2024 (similar to Northern Eurasia pre-2018), we added comparisons between those regionals and World Finals and also included the Europe Championship for reference. That data is kept separate in each table it falls under to separate it from analysis of current superregionals with more than 1 year.

---
2. https://icpc.iisf.or.jp/asia-pacific/2024-2025-cycle/
3. https://nerc.itmo.ru/information/selection-rules.html
4. If received, this paper will be updated with these two regions or removed altogether if not available.

TABLE 1
ICPC regions with historically trackable superregional championships. See Section 1.3 for details.

| Region | Superregional Championship |
| --- | --- |
| Africa and Arab | Africa & Arab Collegiate Programming Championship |
| Asia East | Asia East Continent Final Contest |
| Asia West | Asia West Continent Final Contest |
| North America | North America Championship |
| Northern Eurasia | Northern Eurasia Finals or Northeastern Europe Regional Contest |
| Europe | Europe Championship |
| Southeastern Europe | Southeastern Europe Regional Contest |
| Southwestern Europe | Southwestern Europe Regional Contest |
| Northwestern Europe | Northwestern Europe Regional Contest |
| Central Europe | Central Europe Regional Contest |

For the sake of clarity, when this article refers to "superregionals," it is referring to the championships of the 5 regions in Table 1. To refer to "local regional" contests, we use the term "regional." According to ICPC naming conventions, both are labeled "regional contest" or "regional championship" or some variant of that phrasing, so we separate them for clarity.

## 2 METHODOLOGY

To study the performance consistency across two contests, we propose a methodology based on rank correlation analysis using Kendall's tau coefficient.

For a specific pair of contests, we take the following steps.
1) We first find all pairs of teams that participated in both contests in the same year.
2) We then identify these pairs as being concordant (that is, one team performs better than the other in both contests) or discordant.
3) Finally, we use the formula for Kendall's tau coefficient

$$\tau = \frac{(\text{\# of concordant pairs}) - (\text{\# of discordant pairs})}{(\text{\# of pairs})}.$$

to compute the rank correlation coefficient.

We note that $R^2$ is not directly applicable to analyzing the relationship between ranks. On the other hand, Kendall's Tau rank correlation coefficient is used to assess the strength and direction of monotonic relationships between ranked variables, and is recorded to be more reliable and interpretable than Spearman's rank correlation coefficient, despite being more expensive to compute [22].

Following the cutoffs used in [33], we define the interpretations for values of $\tau$ in Table 2.

### 2.1 World Finals and Superregionals Data Collection

We obtained the rankings of the teams from the ICPC official website [12]. Collecting data from superregional contests posed several challenges due to inconsistencies in reporting formats and the decentralized nature of result dissemination across different superregional websites. In particular, each regional is responsible for collecting and publishing its own contest results, and each region is responsible for the management of regionals, qualification to the superregional contest(s), and overall structure of the region (see Section 1.2). Due to this, many regional results were inaccessible and incomparable to superregional results. For example, in Asia East, all teams are eligible to attend all regionals, which makes correlational data non-independent and more inconsistent. In Sections 3.3 and 3.4, we show results for North America regionals and Northern Eurasia regionals, respectively.

TABLE 2
Interpretations for the magnitude of the Kendall Tau value.
Interpretation *begins* at the listed value, e.g., *Very Strong* is $\geq 0.71$

| Strength | Kendall |
| --- | --- |
| Negligible | 0.00 |
| Weak | 0.06 |
| Moderate | 0.26 |
| Strong | 0.49 |
| Very Strong | $\geq 0.71$ |

### 2.2 Codeforces Data Collection

We obtained the Codeforces[5] rating of teams going to World Finals 2021–2024 from various statistics pages on Codeforces [24], [25], [26] that captured the teams' average rating before the contest. We compare the online rating in these four years to other ICPC contests where the teams overlap to examine the correlation between online contests and ICPC performance. We did not include the European contests as they did not contain much, if any, overlap from 2021–2024 with Codeforces.

## 3 RESULTS

The correlations presented in this section measure if the relative performance of teams from the same superregion in World Finals was the same as their superregional championship. In short, we analyze performance consistency between superregionals and World Finals. We present similar analysis comparing superregionals and World Finals to Codeforces as well as comparing superregionals to regionals for North America and Northern Eurasia.

5. https://codeforces.com/

We limit our analysis to the last 10 years (2015–2024) as data beyond the last decade is less likely to be impactful to present-day decisions. For those interested, all code and results are publicly available on Github at https://github.com/zhtluo/cp-ranking[6]. Adding regions, years, and superregionals is a trivial task if one follows the `readme.md` instructions.

## 3.1 Superregional vs. Final

We gathered the available team rankings from both ICPC World Finals and regions that host superregional finals. The exact range of data is shown in Table 3. We note that Asia East is mostly composed of regionals in China (and North Korea), and therefore did not have superregional contests during COVID (affecting data collection in 2023 and 2024). The Asia East Continent Finals began in 2018, Asia West in 2020[7], and Asia Pacific in 2024. The North America Championship began in 2020. The Europe Championship began in 2024, so that analysis should be read with less weight (only included as a comparison to its regionals in prior years). Northern Eurasia on the official website, icpc.global, does not have records listed on the superregional results contest finder from 2018–2020, but are available under the standard "Northern-Eurasia" tag. See Section 1.3 for more on excluded regions.

TABLE 3
Data range of World Finals and superregional finals used in this study. We used data publicly available on the official website in the past 10 years. For example, there is 10 years of World Finals data available.

| Name | Years |
| --- | --- |
| World Finals | 2015–2024 |
| Africa & Arab Collegiate Programming Championship | 2015–2024 |
| Asia East Continent Final Contest | 2018–2022 |
| Asia West Continent Final Contest | 2020, 2023–2024 |
| North America Championship | 2020–2024 |
| Northern Eurasia Finals | 2015–2024 |
| Europe Championship | 2024 |
| Southeastern Europe Regional Contest | 2017–2023 |
| Southwestern Europe Regional Contest | 2015–2023 |
| Northwestern Europe Regional Contest | 2015–2023 |
| Central Europe Regional Contest | 2015–2023 |

We compute the number of pairs of team overlaps and the respective tau coefficient and list them in Table 4.

We found that overall, team performance in superregional is moderately correlated with team performance in World Finals with coefficient $\tau = 0.407$ (not including the European regions). While most contests hover around the average, team performance in Northern Eurasia gives a particularly strong correlation at $\tau = 0.521$. Moreover, we found that the European regions were also moderately correlated at $\tau = 0.369$ as a whole. Interestingly, Southwestern Europe regionals were not correlated and Central Europe regionals were strongly correlated, indicating a wide range of contest variety across Europe and suggesting further studies into this dynamic superregion.

6. We only request our original work and this paper get referenced when using this repository.
7. Prior to 2020, Asia West did not have superregionals and the ICPC tag used for the Asia West superregionals through present day ("ICPCKolkataKanpur") was used for the Kolkata-Kanpur regionals. Additionally, the Asia West Continent Final data is missing from the official website for the 2021 and 2022 contests, so that data is excluded.

TABLE 4
Rank correlation between superregional and World Finals. We include the number of pairs of teams we used in the calculation and the rank correlation coefficient, rounded to .001. European contests kept separate due to recent (2024) formation of Europe Championship (see Section 1.3).

| Name | # | Coeff. |
| --- | --- | --- |
| Africa & Arab Collegiate Programming Championship | 563 | 0.410 |
| Asia East Continent Final Contest | 453 | 0.395 |
| Asia West Continent Final Contest | 239 | 0.188 |
| North America Championship | 587 | 0.281 |
| Northern Eurasia Finals | 1131 | 0.521 |
| **Weighted Average** | - | **0.407** |
| Europe Championship | 153 | 0.346 |
| Central Europe Regional Contest | 52 | 0.577 |
| Southeastern Europe Regional Contest | 52 | 0.385 |
| Southwestern Europe Regional Contest | 38 | 0.053 |
| Northwestern Europe Regional Contest | 61 | 0.443 |
| **Weighted Average (Incl. EUC)** | - | **0.369** |

## 3.2 Online Rating vs. ICPC

We also measured the rank correlation between teams' average Codeforces online rating and their ranking in superregional finals and World Finals in Table 5.

TABLE 5
Rank correlation between Codeforces and ICPC contests. We include the number of pairs of teams we used in the calculation and the rank correlation coefficient, rounded to .001. We use the overlapping part of Codeforces rating data from 2021–2024 and the superregional data from the table listed in Table 3.

| Name | # | Coeff. |
| --- | --- | --- |
| **World Finals** | **16596** | **0.596** |
| Africa & Arab Collegiate Programming Championship | 130 | 0.415 |
| Asia East Continent Final Contest | 106 | 0.226 |
| Asia West Continent Final Contest | 183 | 0.344 |
| North America Championship | 268 | 0.396 |
| Northern Eurasia Finals | 365 | 0.545 |

We found that Codeforces ratings correlate strongly with World Finals ($\tau = 0.596$), compared to the weighted average of superregional contests with WFs ($\tau = 0.407$). Northern Eurasia Finals ($\tau = 0.545$) also leads other contests by a significant margin.

## 3.3 North America Superregionals

Similarly, we performed comparison between different North America regionals and the North America Championship. We list the data source in Table 6 and the result in Table 7. The table shows $\tau$ with the number of pairwise comparisons used in its calculation. For example, comparing to North American Championships for 2020–2024, East Central had $\{5, 10, 5, 5, 6\}$ teams in each year (only the top team from each school is used, as usually only one per school is allowed to attend superregional championships), from which we take $\binom{n}{2}$ for each year, yielding $\{10, 45, 10, 10, 15\}$ pairs, for a total of 90 pairs. As expected, only teams who attended the superregional championships were counted.

Based on the result, we observe that team performance is on average much more consistent in North America ($\tau = 0.577$) than between North America and World Finals ($\tau = 0.281$).

6TABLE 6
Data range of North America regionals and the North America Championship (superregional contest). We used data publicly available on the official website in the past 10 years. Mid Atlantic and Northeastern are both missing data from 2024 on the website, so those two points are excluded.

| Name | Years |
| --- | --- |
| North America Championship | 2020–2024 |
| East Central | 2015–2024 |
| Mid Atlantic | 2015–2023 |
| Mid Central | 2015–2024 |
| North Central | 2015–2024 |
| Pacific Northwest | 2015–2024 |
| Rocky Mountain | 2015–2024 |
| Southeast | 2015–2024 |
| South Central | 2015–2024 |
| Southern California | 2015–2024 |
| Greater New York | 2015–2024 |
| Northeastern | 2015–2023 |

TABLE 7
Rank correlation between North America regionals and superregional. We include the number of pairs of teams we used in the calculation and the rank correlation coefficient, rounded to .001. Championship data is not available before 2020, so these correlations are measured over data from 2020–2024.

| Name | # | Coefficient |
| --- | --- | --- |
| East Central | 90 | 0.556 |
| Mid Atlantic | 56 | 0.786 |
| Mid Central | 42 | 0.667 |
| North Central | 52 | 0.692 |
| Pacific Northwest | 55 | 0.491 |
| Rocky Mountain | 22 | 0.636 |
| South Central | 30 | 0.467 |
| Southern California | 32 | 0.438 |
| Southeast | 35 | 0.771 |
| Greater New York | 46 | 0.174 |
| Northeastern | 41 | 0.659 |
| **Weighted Average** | - | **0.577** |

## 3.4 Northern Eurasia Superregionals

Similarly, we performed comparison between different Northern Eurasia regionals and the Northern Eurasia Finals. We list the data source in Table 8 and the result in Table 9. There were data gaps around the 2021–2022 Northern Eurasia finals due to some regionals not sending any teams to the Northern Eurasia superregional finals[8].

Furthermore, due to data inconsistency with the official website, at most 6 regions are missing data during the 2018-19 and 2020-21 years. The first was right after Northern Eurasia became its own Region (in 2018), the second was the start of the COVID-19 pandemic, although the contests still happened. This data is excluded from our analysis, although the strength and consistency of Northern Eurasia regions through the other 8 years implies a similar result for the missing data.

Based on the result, we observe that team performance is on average more consistent in Northern Eurasia ($\tau = 0.666$) than between Northern Eurasia and World Finals ($\tau =$

---

8. Northern Eurasia superregional finals were on 4/12/22, lack of attendance (a 70 team drop compared to the average) was likely due to geopolitical events involving war beginning two months prior. A separate "Northern Eurasia Finals: South Caucasus Championship" was held with 21 teams for only this year and was excluded from analysis.

---

TABLE 8
Data range of Northern Eurasia regionals and the Northern Eurasia Finals (superregional contest). We used data publicly available on the official website in the past 10 years. Missing data is listed as [range] \ missing, years. See Section 3.4 for details.

| Name | Years |
| --- | --- |
| Northern Eurasia Finals | 2018–2024 |
| Armenia | 2015–2024 |
| Azerbaijan | 2015–2024 \ 2022[8] |
| Belarus | 2015–2024 |
| Central | 2015–2024 \ 2021 |
| East Siberian | 2015–2024 \ 2019, 21 |
| Far Eastern | 2015–2024 \ 2019, 21 |
| Georgia | 2015–2024 \ 2022[8] |
| Kazakhstan | 2015–2024 \ 2019, 21 |
| Kyrgyzstan | 2015–2024 \ 2019, 21 |
| Moscow | 2015–2024 |
| Northwestern | 2015–2024 |
| Southern and Volga | 2015–2024 \ 2019, 21 |
| Taurida | 2015, 2016, 2020 |
| Tajikistan | 2024 |
| Ural/Eastern | 2015–2024 |
| Uzbekistan | 2015–2024 \ 2017 |
| West Siberian | 2015–2024 |

TABLE 9
Rank correlation between Northern Eurasia regionals and superregional. We include the number of pairs of teams we used in the calculation and the rank correlation coefficient, rounded to .001. Championship data is not available before 2018, so these correlations are measured over data from 2018–2024.
Excluded data (required to have > 1 team advancing to analyze): Tajikistan only had 1 team advance during 2024. Uzbekistan only had 1 team advance during 2017. Tauridia only had > 1 team advance during 2015, 16, and 20.

| Name | # | Coefficient |
| --- | --- | --- |
| Armenia | 57 | 0.754 |
| Azerbaijan | 118 | 0.610 |
| Belarus | 327 | 0.700 |
| Central | 174 | 0.494 |
| East Siberian | 121 | 0.603 |
| Far Eastern | 31 | 0.613 |
| Georgia | 215 | 0.647 |
| Kazakhstan | 296 | 0.757 |
| Kyrgyzstan | 95 | 0.663 |
| Moscow | 345 | 0.659 |
| Northwestern | 259 | 0.807 |
| Southern and Volga | 547 | 0.631 |
| Taurida | 3 | -0.333 |
| Ural/Eastern | 539 | 0.663 |
| Uzbekistan | 304 | 0.638 |
| West Siberian | 349 | 0.679 |
| **Weighted Average** | - | **0.666** |

0.521). Northern Eurasia is only slightly more internally consistent than North America ($\tau = 0.577$), as both are strongly correlated.

## 4 Discussion

**RQ1. Does team performance consistency differ between different ICPC regions? If so, what is the difference?** During our analysis, we found that team performance in **Northern Eurasia** correlates to the World Finals at a significantly stronger scale than other contests (more then .1 difference to the average). We give a visualized representation in Figure 2. This finding corroborates with the folklore understanding that both Russian contests and ICPC

World Finals has a tendency to focus on the observation of problems and the ability to come up with ad-hoc solutions.

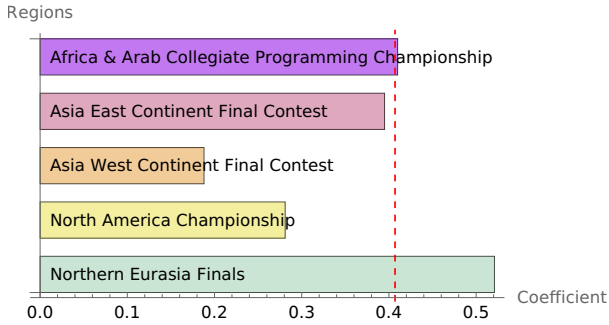

Fig. 2. Rank correlation between superregional and World Finals. Red dashed line represents the average.

**RQ2. Is online competitive programming rating a good indicator of ICPC performance? If so, does it resemble contests from a specific region more than the others?** Based on our results, we see that both World Finals and Northern Eurasia Finals have a strong correlation with Codeforces rating. This result further reinforces the idea that the problem setting in these contests is more similar to online contests, where the ability to do observation and come up with ad-hoc solutions is more important. We illustrate this in Figure 3.

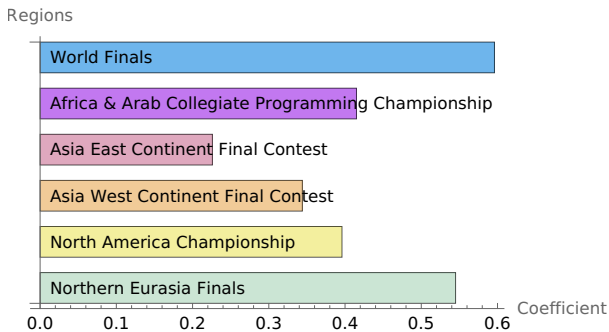

Fig. 3. Rank correlation between Codeforces ratings and ICPC contests.

**RQ3. Can we leverage our understanding of performance consistency to improve competitive programming events?** We show all correlations above 0.35 in Figure 4. We observe that the correlations seem to be clustered — World Finals, Codeforces and Northern Eurasia Finals align with coefficient greater than $\tau = 0.52$ (strongly correlated), while North America Championship, Africa & Arab Collegiate Programming Championship and Asia East Continent Final Contest align with World Finals and/or Codeforces with coefficient close to 0.4.

In addition to later discussion, we suggest that in order to host contests more aligned with World Finals, it could be helpful to study the problem styles and organizations of Northern Eurasia Finals and Codeforces.

**Educational Implications.** Our rank-correlation analysis shows that contests with high tier-to-tier consistency — most notably Northern Eurasia Finals (regional → superregional $\tau \approx 0.67$; superregional → World Finals $\tau \approx 0.52$) — may serve as reliable, low-stakes proxies for eventual WF performance. In such regions, a team's regional standing provides an early, data-driven forecast of its world-level prospects. Instructors can thereby diagnose weaknesses, prescribe targeted practice, and retest in the next regional cycle with reasonable confidence that any rank change reflects genuine skill development rather than contest idiosyncrasies. Conversely, where consistency is weak (e.g., Asia West $\tau \approx 0.19$), regional feedback is noisier, and teams may benefit more from alternative benchmarks such as high-frequency Codeforces rounds (WF $\tau \approx 0.60$). Although we did not measure learning gains directly, these findings suggest that aligning regional problem style with superregional and WF norms could amplify the formative value of regional contests and streamline training pipelines.

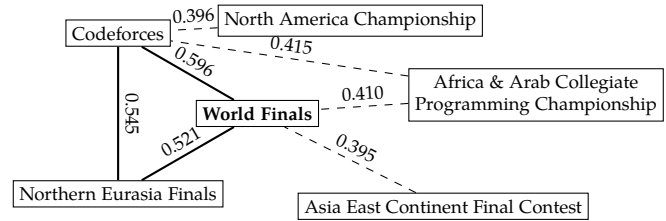

Fig. 4. Team performance correlation of different contests. We list all correlations with Tau coefficient greater than 0.35. Correlations with Tau coefficient greater than 0.49 (strongly correlated) are bolded.

**Practical Implications.** On a practical note, while the correlational coefficients shed some light on the similarities between regionals, superregionals, and worlds, we think that there are more points to consider when making decisions about trainings and contests. For example, ECNA (our region), is strongly correlated with NAC. However, we know anecdotally that ECNA tends to emphasize implementation problems more than NAC does, and we prepare our teams accordingly, despite the strong correlation with NAC ($\tau \approx 0.56$).

Therefore, we recommend considering several other factors, including difficulty of other teams present (many high-performing teams in a region may cause a low correlation with the superregional due to top teams from that region "fighting for first," leading to pseudo-random permutations of those teams), general problem style (hard to quantify), and how well the region performs as a whole at superregionals (e.g. if the entire region performs poorly at superregionals, that may be an indication that regional that those teams spent most of their time preparing for does not align well in structure or style with the superregionals; this could be measured by average or median rank of teams from each region in superregionals). These observations highlight the need for quantitative models of problem style and regional competitiveness, directions we outline next.

## 5 Future Directions and Open Problems

The present work provides an initial, correlational snapshot of performance consistency across ICPC contests, yet several substantive questions remain unanswered. Below we outline the most pressing next steps, both empirical and methodological, that would deepen our understanding of how contest structure and problem design shape team outcomes.

## 5.1 Richer Descriptive Statistics on Regional Performance

Our analyses relied primarily on Kendall's $\tau$ between rank-ordered lists. To capture distributional effects, future studies should report the mean, median, and standard deviation of each superregion's World Finals placings over the last decade and visualize cumulative-density curves of those ranks (one curve per region, five superregions per figure). Such plots will reveal whether a superregion's teams cluster tightly (high internal parity) or exhibit long performance "tails," complementing the pair-wise correlation picture we provide here. The code framework already aggregates contest standings, so adding these statistics is principally an engineering task.

## 5.2 Quantifying Problem-Style Alignment

Our interpretation that Northern Eurasia's problems resemble World Finals and Codeforces sets remains anecdotal. A definitive test requires coding every problem from the last ten years of superregional contests and extracting style features — e.g., topic tags, required algorithms, and solution conciseness. Natural-language processing on statements, plus static analysis of reference solutions, could yield a feature vector per problem. A classifier trained on World Finals vs. other sources would then let us score each regional set for "WF-likeness," producing a quantitative bridge between RQ1 (performance consistency) and RQ3 (design recommendations). Because problem material is often private, securing cooperation from ICPC and regional organizers is a prerequisite.

## 5.3 Repeated or Multi-Stage Superregionals

Codeforces outperforms every superregional as a predictor of WF ranks ($\tau = 0.596$ vs. 0.407 superregional average) largely because it offers hundreds of rating samples per year. A natural experiment is running two superregional rounds (e.g., fall and spring) and testing whether the second round improves the correlation with World Finals. This design would also let us measure within-team variance across similar, high-stakes contests, an aspect invisible in one-shot championships. Unfortunately, such a fundamental change to contest structure is unlikely at the superregional level, but may be viable at a regional or subregional level.

## 5.4 Discrimination and Resolution of Problem Sets

High correlation does not guarantee that a contest discriminates effectively among world-class teams. For each superregional, we propose computing a discrimination factor: the ratio of pairwise rank inversions among advancing teams when moving from superregional to World Finals. Low ratios indicate that the earlier contest already separated teams on the relevant dimensions of skill; high ratios suggest that the superregional set failed to expose meaningful differences. A similar metric can be calculated between regionals and superregionals.

## 5.5 Cross-Circuit Generalization

While this study focused on Codeforces, extending the dataset to AtCoder, LeetCode, and HackerRank ratings would test whether the predictive power stems from platform diversity or contest frequency. A multivariate regression on multiple online ratings could isolate which platform characteristics (problem style, duration, scoring, editorial quality) explain additional variance in WF performance beyond Codeforces alone. We conjecture that Codeforces is the top-performing of these datasets, as it is widely regarded as the most authoritative online-only source of CP rankings.

## 5.6 Longitudinal Team Tracking and Causal Inference

Our correlation design cannot disentangle selection effects from causal ones. Building a panel dataset that follows individual students across several seasons (recording training regimen, online-contest volume, and team composition changes) would enable difference-in-differences or hierarchical-mixed modeling. Such designs could test, for instance, whether adding a Codeforces-style "observation" problem to a regional directly improves later WF rank for the same team.

## 5.7 Data Completeness and Imputation

World finalists were the primary target of [24], [25], [26], but most regions lose a substantial number of observations at the regional level. Before extending the analysis, we must document the exact exclusion counts and test multiple imputation strategies (e.g., Bayesian ridge, k-nearest neighbors) to ensure that missing data do not bias $\tau$-estimates downward. A sensitivity analysis should accompany any imputed results. Furthermore, deep investigation has yielded two new Codeforces datasets from 2020 and 2019[9], which should be downloaded, formatted, and added to present analysis.

## 5.8 Comparative Case Studies Beyond ICPC

Preliminary checks show China's IOI pipeline (NOIP→NOI) yields $\tau \approx 0.3773$ — remarkably close to the averages we observe for ICPC superregionals. Thoroughly analyzing such parallel ecosystems will test the external validity of our claims and may uncover structural elements (e.g., national training camps, centralized problem-setting committees) that generalize across competitions.

## 5.9 Educational Impact Metrics

Finally, to honor the educational motivation of competitive programming, future work should measure learning gains — not just rank correlations. Embedding short conceptual quizzes before and after training phases, or analyzing code-quality metrics in post-contest repositories, would reveal whether regions with higher performance consistency also foster deeper algorithmic understanding.

---

9. https://codeforces.com/blog/Laggy

# 6 Conclusion

Addressing the directions listed above will transform the present exploratory correlations into a comprehensive, evidence-driven model of how contest architecture influences educational and competitive outcomes. We invite collaboration from contest organizers, educators, and data owners to make these next steps possible and to move competitive programming research from anecdote toward reproducible science.

## Acknowledgments

The authors thank Christian Lim for support in connecting the authors with various experts around the world on particular regions. The authors also thank the following people for aiding in the detailed description of each region's structure and qualifications: Mohamed Fouad (Africa and Arab), Steven Halim (Asia Pacific), Benjamin Rubio and Federico Meza (Latin America), and Lidia Perovskaya (Northern Eurasia).